\documentclass[aip,pof,longbibliography,twocolumn,reprint]{revtex4-1}
\usepackage[utf8]{inputenc}
\usepackage{natbib}
\usepackage{amssymb,amsmath}
\usepackage{epsfig}
\usepackage{bm}
\usepackage{color}

\usepackage{graphicx}% Include figure files

\usepackage{color}
\newcommand{\vicente}[1]{{ #1}}
\newcommand{\vicenteg}[1]{{ #1}}
\newcommand\beq{\begin{equation}}
\newcommand\eeq{\end{equation}}
\newcommand\beqa{\begin{eqnarray}}
\newcommand\eeqa{\end{eqnarray}}

\newcommand{\al}{\alpha}

\begin{document}
\title{Applications of the kinetic theory for a model of a confined quasi-two dimensional granular mixture: Stability analysis and thermal diffusion segregation}

%\author{Rub\'en G\'omez Gonz\'alez\footnote[1]{Electronic address: ruben@unex.es}}
%\affiliation{Departamento de F\'{\i}sica,
%Universidad de Extremadura, E-06006 Badajoz, Spain}
%\author{Nagi Khalil\footnote[2]{Electronic address: nagi@ifisc.uib-csic.es}}
%\affiliation{IFISC (CSIC-UIB), Instituto de F\'{\i}sica Interdisciplinar y Sistemas Complejos,
%Campus Universitat de les Illes Balears,
%E-07122, Palma de Mallorca, Spain}
\author{Vicente Garz\'o}
\email{vicenteg@unex.es} \homepage{http:https://fisteor.cms.unex.es/investigadores/vicente-garzo-puertos/}
\affiliation{Departamento de
F\'{\i}sica and Instituto de Computaci\'on Cient\'{\i}fica Avanzada (ICCAEx), Universidad de Extremadura, E-06071 Badajoz, Spain}
\author{Ricardo Brito}
\affiliation{Departamento de Estructura de la Materia, F\'{\i}sica T\'ermica y Electr\'onica and GISC, Universidad Complutense de Madrid, E-28040 Madrid, Spain}
\author{Rodrigo Soto}
\affiliation{Departamento de
F\'{\i}sica, Facultad de Ciencias F\'{\i}sicas y Matem\'aticas, Universidad de Chile, 8370449 Santiago, Chile}

\begin{abstract}

The Boltzmann kinetic theory for a model of a confined quasi-two dimensional granular mixture derived previously [Garz\'o, Brito and Soto, Phys. Fluids \textbf{33}, 023310 (2021)] is considered further to analyze two different problems. First, a linear stability analysis of the hydrodynamic equations with respect to the homogeneous steady state (HSS) is carried out to identify the conditions for stability as functions of the wave vector, the coefficients of restitution, and the parameters of the mixture. The analysis, which is based on the results obtained by solving the Boltzmann equation by means of the Chapman--Enskog method to first order in spatial gradients, takes into account the (nonlinear) dependence of the transport coefficients and the cooling rate on the coefficients of restitution and applies in principle to arbitrary values of the concentration, and the mass and diameter ratios. In contrast to the results obtained in the conventional inelastic hard sphere (IHS) model, the results show that all the hydrodynamic modes are stable so that,
the HSS is linearly \emph{stable} with respect to long enough wavelength excitations. On the other hand, this conclusion agrees with previous stability analysis performed in earlier studies for monocomponent granular gases. As a second application, segregation induced by both a thermal gradient and gravity is studied. A segregation criterion based on the dependence of the thermal diffusion factor $\Lambda$ on the parameter space of the mixture is derived. \vicenteg{Our results} show the transition between $\Lambda>0$ (larger particles tend to move toward the cold plate) to $\Lambda<0$ (larger particles tend to move toward the hot plate) by varying the parameters of the system (masses, sizes, composition and coefficients of restitution). Comparison with previous results derived from the IHS model is carried out.

\end{abstract}

\maketitle

\section{Introduction}
\label{sec1}

When granular matter is externally excited, the motion of grains is quite similar to the chaotic motion of atoms or molecules of an ordinary fluid. In this situation (rapid-flow conditions), they admit a hydrodynamic-like type of description which can be derived from a fundamental point of view by using kinetic theory tools. \cite{BP04,G19} However, kinetic theory must be adapted to dissipative dynamics since the dimension of grains is macroscopic (typically of micrometers or larger), and hence their collisions are inelastic. As a consequence, to keep them in rapid-flow conditions, one needs to subject to the grains to  a violent and sustained excitation to compensate for the energy lost by collisions and achieve a non-equilibrium steady state.     

In real experiments, there are several ways of injecting energy into the system. For instance, it can be done by vibrating the walls of the system, \cite{YHCMW02,HYCMW04} or by bulk driving (as in air-fluidized beds). \cite{AD06,SGS05} Nevertheless, it is well-known that this sort of heating produces in general strong spatial gradients whose theoretical description goes beyond the Navier--Stokes hydrodynamic equations (which hold for small spatial gradients). Thus, to avoid the difficulties embodied in the theoretical description of far from equilibrium situations, it is quite usual to drive the granular gas by means of external forces or \emph{thermostats}. \cite{PLMV99,CLH00,PEU02,PTNE02,PBL02,FAZ09,VAZ11,GSVP11} However, unfortunately it is not quite clear the relationship between the results obtained with thermostats with those provided in the real experiments.

A possible way of avoiding the use of external driving forces has been proposed in the granular literature in the past few years. \cite{OU98,PMEU04,MVPRKEU05,CCDHMRV08,PCR09,RPGRSCM11,CMS12} The idea is to consider a particular geometry (quasi-two dimensional geometry) where the granular gas is confined in the vertical direction of the box; this direction is slightly larger than one particle diameter. Thus, when the box is vertically vibrated, energy is supplied into the vertical degrees of freedom of grains through the collisions of them with the top and bottom plates. The energy lost by collisions is counterbalanced by the energy gained by grains due to their collisions with the walls. This energy is transferred to the horizontal degrees of freedom of grains. Under these conditions, when the system is observed from above, it is fluidized and can remain in a homogeneous state.     

Needless to say, the collisional dynamics of the above geometry is quite intricate due essentially to the drastic restrictions imposed by the the confinement to the set of possible impact parameters. \cite{MGB19a} Thus, although some attempts have been recently made \cite{MBGM22,MGB22,MPGM23} by considering explicitly the confinement in the Boltzmann collision operator, a collisional model was proposed years ago by Brito \emph{et al.} \cite{BRS13} to gain some insight into this quite complex problem. In this model, the magnitude of the normal component of the relative velocity of the colliding spheres is increased by a factor $\Delta$ in each collision. The term associated with the factor $\Delta$ in the scattering rule attempts to mimic the transfer of energy from the vertical degrees of freedom of grains to the horizontal ones.

The collisional model with constant $\Delta$ (henceforth, it will be referred to here as the $\Delta$-model) has been widely considered by different researchers in the past few years to analyze the dynamic properties of granular gases in the quasi-two dimensional geometry. In particular, for monocomponent dilute granular gases, the $\Delta$-model has been considered for studying the homogeneous state, \cite{BGMB13,BMGB14} for deriving expressions of the Navier--Stokes transport coefficients, \cite{BBMG15} and for performing a linear stability analysis of the homogeneous time-dependent state. \cite{BBGM16} Independently, Soto \emph{et al.} \cite{SRB14} have determined the shear viscosity coefficient of a dilute granular gas; their theoretical results compare very well with computer simulations. The above works have been extended then \cite{GBS18,GBS20,GBS21a} to moderate densities by considering the inelastic version of the Enskog kinetic equation.

More recently, the $\Delta$-model has been extended to binary mixtures in the low-density regime in two different papers. First, the homogeneous steady state (HSS) has been widely studied in Ref.\ \onlinecite{BSG20} with special emphasis on the breakdown of energy equipartition. Then, the (inelastic) Boltzmann equation has been solved by means of the Chapman-- Enskog perturbative method \cite{CC70} for \vicente{states near the local version of the homogeneous time-dependent state. Since the zeroth-order velocity distribution function (reference state of the Chapman--Enskog method) $f_i^{(0)}(\mathbf{r}, \mathbf{r}; t)$ of the species $i$ is a normal solution, its dependence on time only occurs through the (global) granular temperature.} In the first order of the expansion, explicit expressions of the set of Navier--Stokes transport coefficients have been derived by \vicente{assuming the steady state conditions (HSS) and by considering the leading terms in a Sonine polynomial expansion. \cite{GBS21} These expressions were derived by considering the case $\Delta_{11}=\Delta_{22}=\Delta_{12}$, where $\Delta_{ij}$ characterizes the intensity of the energy injection term associated with collisions between particles of the species $i$ and $j$.}

The knowledge of the Navier--Stokes transport coefficients opens up the possibility of studying different problems. Among them, the analysis of the stability of the HSS is not only an interesting application of the Navier--Stokes hydrodynamic equations by itself, but also because the \vicente{transport coefficients of the mixture have been evaluated in this steady state. In this context, the HSS plays a similar role}
as the so-called homogeneous cooling state (HCS) does in the conventional inelastic hard sphere (IHS) model. In this latter case, it is well known
\cite{GZ93,M93} that the HCS becomes unstable when
the linear size of the system, $L$, is larger than a certain critical
length $L_c$. The dependence of the critical length $L_c$ on the parameter space of the system can be obtained from a linear
stability analysis of the Navier--Stokes hydrodynamic equations. Theoretical predictions for $L_c$ \cite{BDKS98,G05,GMD06,G15} have been shown
to compare very well with computer simulations, \cite{BRM98,MDCPH11,MGHEH12,BR13,MGH14}
even for strong inelasticities. This good agreement reinforces the reliability of kinetic theory for describing granular flows.

In contrast to the HCS, the linear stability analysis carried out here shows that the HSS is linearly stable with respect to long-enough wavelength perturbations. This conclusion agrees with
previous stability analysis performed in the $\Delta$-model for monocomponent granular gases at low \cite{BBGM16} and moderate \cite{GBS21a} densities.  However, as expected,
the forms of the $d-1$ transversal shear modes ($d$ being the
dimensionality of the system) and the four longitudinal modes
(i.e., those associated with the mole fraction, the longitudinal component of the flow velocity, the hydrostatic pressure, and the temperature) derived in this paper differ from those previously obtained \cite{GMD06} in the HCS for a granular binary mixture.

The fact that the linearized hydrodynamic equations are linearly stable does not preclude the possibility of separation or species segregation induced by both gravity and thermal gradients. Segregation and mixing of dissimilar grains is one of the most interesting problems in granular mixtures, from a fundamental and a practical point of view. The knowledge of the diffusion transport coefficients allows us to derive a segregation criterion based on the sign of the so-called thermal diffusion factor $\Lambda$. Since in our geometry (see Fig.\ \ref{sketch}) the bottom plate is hotter than the top plate (and so, gravity and thermal gradient point in parallel directions), then when $\Lambda>0$ ($\Lambda<0$) the larger particles tend to move toward the cold (hot) plate. The present study complements previous works made in the context of kinetic theory of IHS, \cite{JY02,BRM05,SGNT06,G06,G08a,G09,G11} and more recently for granular suspensions. \cite{GG23} As expected, our results show that the dependence of the marginal segregation curve ($\Lambda=0$) on the parameter space of the system in the $\Delta$-model differs from the one obtained before for IHS.

The plan of the paper is as follows. In Sec.\ \ref{sec2} the Boltzmann kinetic equation for a granular binary mixture in the $\Delta$-model is introduced. Then, the corresponding Navier--Stokes transport equations for the mixture are displayed and the HSS analyzed. Section \ref{sec3} deals with the (linear) stability analysis of hydrodynamic equations. As in the case of the conventional IHS, \cite{GMD06} the $d-1$ transversal shear modes are decoupled from the four longitudinal hydrodynamic modes. Since the shear viscosity coefficient is always positive, then the transversal shear modes are linearly stable. However, the study of the evolution of the longitudinal hydrodynamic
modes is much more intricate and requires in general to resort to a numerical analysis. For this reason, the \vicente{extreme long wavelength limit (wave vector k = 0)} is
previously studied; the analysis shows that the longitudinal modes are also linearly stable. For nonzero values of the wave vector
(which is equivalent to consider the terms coming from the
spatial gradients in the constitutive equations), a systematic analysis of the dependence of the
longitudinal modes on the control parameters shows that these
modes also decay in time and so we can conclude that the HSS is linearly stable. Thermal diffusion segregation is studied in Sec.\ \ref{sec4} while the paper is closed in Sec.\ \ref{sec5} with a brief discussion of the results reported here.

\section{Boltzmann kinetic equation for granular binary mixtures}
\label{sec2}

\subsection{Model of a confined quasi-two dimensional granular mixture}

We consider a granular binary mixture modeled as a gas of inelastic hard spheres of masses $m_1$ and $m_2$ and diameters $\sigma_1$ and $\sigma_2$. We assume that the spheres are completely \emph{smooth} and hence, the inelasticity in collisions is fully characterized by the constant (positive) coefficients of normal restitution $\al_{ij}\leq 1$ for collisions between particles of species (or components) $i$ and $j$. The granular gas is in the presence of a gravitational field ${\bf g}=-g \hat{{\bf e}}_z$, where $g$ is a positive constant and $\hat{{\bf e}}_z$ is the unit vector in the positive direction of the $z$ axis. In the low-density regime, the velocity distribution function $f_i(\mathbf{r}, \mathbf{v};t)$ of the species $i$ ($i=1,2$) in the $\Delta$-model obeys the Boltzmann kinetic equation
\beq
\label{1.1}
\frac{\partial}{\partial t}f_i+\mathbf{v}\cdot \nabla f_i+\mathbf{g}\cdot \frac{\partial f_i}{\partial \mathbf{v}}=\sum_{j=1}^2\; J_{ij}[\mathbf{r},\mathbf{v}|f_i,f_j],
\eeq
where the Boltzmann collision operators $J_{ij}$ of the $\Delta$-model read \cite{BSG20}
\beqa
\label{1.2}
& & J_{ij}[\mathbf{v}_1|f_i,f_j]\equiv \sigma_{ij}^{d-1} \int d{\bf v}_{2}\int d \widehat{\boldsymbol{\sigma}}\;
\Theta (-\widehat{{\boldsymbol {\sigma }}}\cdot {\bf g}_{12}-2\Delta_{ij})\nonumber\\
& & \times
(-\widehat{\boldsymbol {\sigma }}\cdot {\bf g}_{12}-2\Delta_{ij})
\al_{ij}^{-2}f_i(\mathbf{r},\mathbf{v}_1'';t)f_j(\mathbf{r},\mathbf{v}_2'';t)\nonumber\\
& & -\sigma_{ij}^{d-1}\int d {\bf v}_{2}\int d\widehat{\boldsymbol{\sigma}}\;
\Theta (\widehat{{\boldsymbol {\sigma }}}\cdot {\bf g}_{12})
(\widehat{\boldsymbol {\sigma }}\cdot {\bf g}_{12})f_i(\mathbf{r},\mathbf{v}_1;t)\nonumber\\
& &\times
f_j(\mathbf{r},\mathbf{v}_2;t),
\eeqa
where $\Theta(x)$ is the Heaviside step function, $\mathbf{g}_{12}=\mathbf{v}_1-\mathbf{v}_2$ is the relative velocity, $\widehat{{\boldsymbol {\sigma}}}$ is the unit collision vector joining the centers of the two colliding spheres and pointing from particle 1 to particle 2, $\boldsymbol{\sigma}_{ij}=\sigma_{ij}\widehat{\boldsymbol{\sigma}}$ and $\sigma_{ij}=(\sigma_i+\sigma_j)/2$. In Eq.\ \eqref{1.2}, note that particles are approaching if $\widehat{{\boldsymbol {\sigma}}}\cdot \mathbf{g}_{12}>0$. The relationship between the pre-collision velocities $(\mathbf{v}_1'',\mathbf{v}_2'')$
and the post-collision velocities $(\mathbf{v}_1,\mathbf{v}_2)$ is given by
\beq
\label{1.3}
\mathbf{v}_1''=\mathbf{v}_1-\mu_{ji}\left(1+\alpha_{ij}^{-1}\right)(\widehat{{\boldsymbol {\sigma }}}\cdot \mathbf{g}_{12})\widehat{{\boldsymbol {\sigma }}}-2\mu_{ji}\Delta_{ij}\al_{ij}^{-1} \widehat{{\boldsymbol {\sigma }}},
\eeq
\beq
\label{1.4}
\mathbf{v}_2''=\mathbf{v}_2+\mu_{ij}\left(1+\alpha_{ij}^{-1}\right)(\widehat{{\boldsymbol {\sigma }}}\cdot \mathbf{g}_{12})\widehat{{\boldsymbol {\sigma }}}+2\mu_{ij}\Delta_{ij}\al_{ij}^{-1} \widehat{{\boldsymbol {\sigma }}},
\eeq
where $\mu_{ij}=m_i/(m_i+m_j)$. The quantity $\Delta_{ij}$ (which points outward in the normal direction $\widehat{{\boldsymbol {\sigma }}}$ as required by the conservation of angular momentum \cite{L04bis}) is an extra velocity added to the relative motion. In addition, it must be recalled that although the $\Delta$-model has been built to describe quasi-two dimensional systems, the calculations worked out here will be performed for an arbitrary number of dimensions $d$.

In the case of a binary granular mixture, the relevant hydrodynamic fields are the number densities
\beq
\label{1.5}
n_{i}(\mathbf{r};t)=\int d{\bf v}\; f_{i}(\mathbf{r}, \mathbf{v};t),
\eeq
the flow velocity $\mathbf{U}$
\beq
\label{1.6}
\mathbf{U}(\mathbf{r};t)=\rho(\mathbf{r};t)^{-1}\sum_{i=1}^2\; m_{i}
\int d {\bf v}\; {\bf v}\; f_{i}(\mathbf{r}, \mathbf{v};t),
\eeq
and the granular temperature $T$
\beq
\label{1.7}
T(\mathbf{r};t)=\frac{1}{n(\mathbf{r};t)}\sum_{i=1}^2\frac{m_{i}}{d}\int d{\bf
v}\; V^{2}\; f_{i}(\mathbf{r}, \mathbf{v};t).
\eeq
Here, ${\bf V}={\bf v}-{\bf U}$ is the peculiar velocity, $n=n_{1}+n_{2}$ is the total number density, and $\rho
=m_{1}n_{1}+m_{2}n_{2}$ is the total mass density. Apart from the hydrodynamic fields, an interesting quantity is the partial temperature $T_i$. This quantity measures the mean kinetic energy of species $i$. It is defined as
\beq
\label{1.9}
T_i(\mathbf{r};t)=\frac{m_i}{d n_i(\mathbf{r};t)}\int d{\bf v}\; V^{2}\; f_{i}(\mathbf{r}, \mathbf{v};t).
\eeq

\subsection{Hydrodynamic equations}

The corresponding (macroscopic) hydrodynamic equations for the number densities $n_i(\mathbf{r};t)$, the flow velocity $\mathbf{U}(\mathbf{r};t)$, and the granular temperature $T(\mathbf{r};t)$ can be derived from the set of Boltzmann kinetic equations \eqref{1.1} when one takes into account that the operators $J_{ij}[f_i,f_j]$ conserve the particle number of each species, the total momentum but the total energy is not conserved due to the inelasticity of collisions. The structure of the hydrodynamic equations is similar to that of the conventional IHS model, \cite{G19} and they are given by
\begin{equation}
D_{t}n_{i}+n_{i}\nabla \cdot {\bf U}+\frac{\nabla \cdot {\bf j}_{i}}{m_{i}}
=0,  \label{1.10}
\end{equation}
\begin{equation}
D_{t}{\bf U}+\rho ^{-1}\nabla \cdot \mathsf{P}-\mathbf{g}=0,  \label{1.11}
\end{equation}
\begin{equation}
D_{t}T-\frac{T}{n}\sum_{i=1}^2\frac{\nabla \cdot {\bf j}_{i}}{m_{i}}+\frac{2}{dn}
\left( \nabla \cdot {\bf q}+\mathsf{P}:\nabla {\bf U}\right) =-\zeta \,T.
\label{1.12}
\end{equation}
In the above equations, $D_{t}=\partial_{t}+{\bf U}\cdot \nabla $ is the
material derivative,
\begin{equation}
{\bf j}_{i}=m_{i}\int d{\bf v}\,{\bf V}\,f_{i}({\bf v}), \quad \mathbf{j}_1=-\mathbf{j}_2
\label{1.13}
\end{equation}
is the mass flux for the component $i$ relative to the local flow,
\begin{equation}
\mathsf{P}=\sum_{i=1}^2\,\int d{\bf v}\,m_{i}{\bf V}{\bf V}\,f_{i}({\bf v})  \label{1.14}
\end{equation}
is the total pressure tensor,
\begin{equation}
{\bf q}=\sum_{i=1}^2\,\int d{\bf v}\,\frac{1}{2}m_{i}V^{2}{\bf V}
\,f_{i}({\bf v})  \label{1.15}
\end{equation}
is the total heat flux, and
\begin{equation}
\zeta=-\frac{2}{d n T}\sum_{i=1}^2\sum_{j=1}^2\int d{\bf v}\frac{1}{2}m_{i}v^{2}J_{ij}[{\bf v}|f_{i},f_{j}]
\label{1.16}
\end{equation}
is the (total) cooling rate due to inelastic collisions among all the species. At a kinetic level, we can introduce the ``cooling rates'' $\zeta_i$ associated with the partial temperatures $T_i$. They are defined as
\beq
\label{1.16.1}
\zeta_i=-\frac{1}{d n_i T_i}\sum_{j=1}^2 \; \int d\mathbf{v}\; m_i v^2\; J_{ij}[\mathbf{v}|f_i,f_j].
\eeq

In the case of a binary mixture, there are $d+3$ independent fields: $n_{1}$, $n_{2}$, $\mathbf{U}$, and $T$. Needless to say, Eqs.\ \eqref{1.10}--\eqref{1.12} do not constitute a closed set of differential equations unless one expresses the fluxes and the cooling rate in terms of the hydrodynamic fields. Such expressions are called ``constitutive equations''. As has been previously discussed in some works, \cite{GD02,GMD06} it is more convenient in the low-density regime to provide these constitutive equations in terms of a different set
of experimentally more accessible fields: the composition (or mole fraction of species 1) $x_{1}=n_1/n$, the hydrostatic pressure 
\beq
\label{1.8}
p=n T=\sum_{i=1}^2 n_i T_i,
\eeq
the flow velocity  $\mathbf{U}$, and the granular temperature $T$. The hydrodynamic equations for $x_1$ and $p$ can be easily obtained from Eqs.\ \eqref{1.10} and \eqref{1.12} as
\begin{equation}
D_{t}x_{1}+\frac{\rho }{n^{2}m_{1}m_{2}}\nabla \cdot \mathbf{j}_{1}=0\;,
\label{1.17}
\end{equation}
\begin{equation}
D_{t}p+p\nabla \cdot \mathbf{U}+\frac{2}{d}\left( \nabla \cdot \mathbf{q}+
\mathsf{P}:\nabla \mathbf{U}\right) =-\zeta p.  \label{1.18}
\end{equation}

In terms of the fields $x_1$, $\mathbf{U}$, $p$, and $T$, the detailed form of the constitutive equations up to the first order in the spatial gradients (Navier--Stokes hydrodynamic order) are given by \cite{GBS21}
\beq
\label{1.19}
{\bf j}_{1}=-\frac{m_{1}m_{2}n}{\rho} D\nabla x_{1}-\frac{\rho}{p}D_{p}\nabla p-\frac{\rho}{T}D_T\nabla T,
\eeq
\beq
\label{1.20}
\vicenteg{P_{\ell k}^{(1)}=-\eta\left(\nabla_k U_\ell+\nabla_\ell U_k-\frac{2}{d}\delta_{\ell k}\nabla \cdot \mathbf{U}\right)},
\eeq
\vspace{0.1cm}
\beq
\label{1.21}
\mathbf{q}^{(1)}=-T^2 D''\nabla x_1-L \nabla p-\lambda \nabla T,
\eeq
\beq
\label{1.22}
\zeta=\zeta^{(0)}+\zeta_U \nabla \cdot \mathbf{U},
\eeq
\vicenteg{where $\nabla_k\equiv \partial/\partial r_k$}. In Eqs.\ \eqref{1.19}--\eqref{1.22}, the Navier--Stokes transport coefficients are the diffusion coefficient $D$, the pressure diffusion coefficient $D_p$, the thermal diffusion coefficient $D_T$, the shear viscosity coefficient $\eta$, the Dufour coefficient $D''$, the pressure energy coefficient $L$, and the thermal conductivity coefficient $\lambda$. Moreover, $\zeta^{(0)}$  is the zeroth-order contribution to the cooling rate and the coefficient $\zeta_U$ vanishes in the conventional IHS for dilute binary mixtures. \cite {GD02} \vicente{It is important to remark that in the Chapman--Enskog method, one has to characterize the magnitude of the gravitational force relative to spatial gradients. In the derivation of Eqs.\ \eqref{1.19}--\eqref{1.22}, it has been assumed that the magnitude of the gravity is at least of first-order in perturbation expansion. This is the usual hypothesis in the conventional Chapman--Enskog method for elastic collisions. \cite{CC70}}

Approximate expressions for all the above transport coefficients were obtained in Ref.\ \onlinecite{GBS21} in the leading Sonine approximation in the case $\Delta_{11}=\Delta_{22}=\Delta_{12}\equiv \Delta$. This is the usual case studied for binary mixtures, namely, when the different species differ in their masses, diameters, composition, and coefficients of restitution. We will consider henceforth this case in the remaining part of the paper.

Substitution of the Navier--Stokes constitutive equations, \eqref{1.19}--\eqref{1.22}, into the exact balance equations \eqref{1.11}, \eqref{1.12}, \eqref{1.17}, and \eqref{1.18} gives the Navier--Stokes hydrodynamic equations for a binary mixture in the $\Delta$-model:
\begin{widetext}
\begin{equation}
D_{t}x_{1}=\frac{\rho }{n^{2}m_{1}m_{2}}\nabla \cdot \left( \frac{
m_{1}m_{2}n }{\rho }D\nabla x_{1}+\frac{\rho }{p}D_{p}\nabla p+\frac{\rho }{T
}D^{\prime }\nabla T\right) \;,  \label{1.23}
\end{equation}
\begin{equation}
D_{t}U_{\ell}+\rho^{-1}\nabla_{\ell}p=\mathbf{g}+\rho^{-1}\nabla_{k}\left[\eta \left(
\nabla_{\ell}U_{k}+\nabla_{k}U_{\ell}-\frac{2}{d}\delta_{k\ell}\nabla
\cdot \mathbf{U}\right)\right] \;,  \label{1.24}
\end{equation}
\begin{eqnarray}
\left(D_{t}+\zeta^{(0)} \right) p+\frac{d+2}{d}p\nabla \cdot \mathbf{U} &=&\frac{2}{
d}\nabla \cdot \left( T^{2}D^{\prime \prime }\nabla x_{1}+L\nabla
p+\lambda\nabla T\right)  \nonumber \\
&&+\frac{2}{d}\eta \left(\nabla_{\ell}U_{k}+\nabla_{k}U_{\ell}-\frac{2}{
d}\delta_{k\ell }\nabla \cdot \mathbf{U}\right) \nabla_{\ell}U_{k}-p \zeta_U \nabla \cdot \mathbf{U},
\label{1.25}
\end{eqnarray}
\begin{eqnarray}
\left(D_{t}+\zeta^{(0)} \right) T+\frac{2}{d}T\nabla \cdot \mathbf{U} &=&-\frac{T
}{n}\frac{m_{2}-m_{1}}{m_{1}m_{2}}\nabla \cdot \left( \frac{m_{1}m_{2}n}{
\rho }D\nabla x_{1}+\frac{\rho }{p}D_{p}\nabla p+\frac{\rho }{T}%
D^{\prime}\nabla T\right)  \nonumber \\
&&+\frac{2}{d n}\nabla \cdot \left( T^{2}D^{\prime \prime }\nabla
x_{1}+L\nabla p+\lambda \nabla T\right)  \nonumber \\
&&+\frac{2}{d n}\eta \left( \nabla_{\ell}U_{k}+\nabla_{k}U_{\ell}-\frac{2
}{d}\delta _{k\ell}\nabla \cdot \mathbf{U}\right) \nabla_{\ell}U_{k}-T \zeta_U \nabla \cdot \mathbf{U}.
\label{1.26}
\end{eqnarray}
\end{widetext}
For the chosen set of fields $n=p/T$ and $\rho =p\left[ \left(
m_{1}-m_{2}\right) x_{1}+m_{2}\right] /T$. Equations \eqref{1.23}--\eqref{1.26} are exact to
second order in the spatial gradients for a low density Boltzmann gas.

\subsection{Homogeneous steady state (HSS)}

For general time-dependent states, the expressions of the transport coefficients and the cooling rate in the hydrodynamic regime can be written as
\beq
\label{1.27}
D=\frac{\rho T}{m_{1}m_{2}\nu}D^*, \quad D_p=\frac{p}{\rho \nu}D_p^*, \quad D_T=\frac{p}{\rho \nu}D_T^*,
\eeq
\beq
\label{1.28}
\eta=\frac{p}{\nu}\eta^*, \quad D''=\frac{n}{\overline{m}\nu}D^{''*}, \quad L=\frac{T}{\overline{m}\nu}L^*,
\eeq
\beq
\label{1.29}
\lambda=\frac{p}{\overline{m}\nu}\lambda^*, \quad \zeta_0^*=\frac{\zeta^{(0)}}{\nu}.
\eeq
Here, $\nu=n \sigma_{12}^{d-1} v_\text{th}$ is an effective collision frequency, $v_\text{th}=\sqrt{2T/\overline{m}}$ is the thermal velocity and $\overline{m}=(m_1+m_2)/2$. The reduced transport coefficients ($D^*$, $D_p^*$, $D_T^*$, $\eta^*$, $D^{''*}$, $L^*$, $\lambda^*$, and $\zeta_0^*$) depend on the concentration $x_1$, the mass and diameter ratios $m_1/m_2$ and $\sigma_1/\sigma_2$, the coefficients of restitution $\al_{ij}$, and the (reduced) velocity $\Delta^*=\Delta/v_\text{th}$. On the other hand, as discussed in previous papers, \cite{BBGM16,BSG20,GBS21} the determination of the (reduced) transport coefficients requires in general to numerically solve intricate first-order differential equations in the dimensionless parameter $\Delta^*(T)$. This contrasts with the results obtained from the conventional IHS, \cite{GD02,GMD06,GM07} where those reduced coefficients are independent of the time-dependent granular temperature $T(t)$.

A simple but quite interesting situation corresponds to the steady state solution characterized by the absence of gravity ($g=0$) and where the pressure and the temperature reach constant values for asymptotically long times. In the HSS, $\partial_t p=\partial_t T=0$ and according to Eqs.\ \eqref{1.25} and \eqref{1.26} the partial cooling rates vanish ($\zeta_1^{(0)}=\zeta_2^{(0)}=\zeta^{(0)}=0$). As Eq.\ \eqref{1.16.1} shows, the quantities $\zeta_i^{(0)}$ are given in terms of the unknown zeroth-order distributions  $f_i^{(0)}$ of the Chapmann--Enskog solution. However, a good estimate on $\zeta_i^{(0)}$ can be obtained by replacing $f_i^{(0)}$ by the Maxwellian distribution defined at the zeroth-order contribution $T_i^{(0)}$ to the partial temperature $T_i$:
\beq
\label{1.29.1}
f_i^{(0)}(\mathbf{V})\to f_{i,\text{M}}(\mathbf{V})=n_i \left(\frac{m_i}{2\pi T_i^{(0)}}\right)^{d/2}\exp \left(-\frac{m_i V^2}{2 T_i^{(0)}}\right).
\eeq
By using the Maxwellian distributions \eqref{1.29.1}, the (reduced) partial cooling rates $\zeta_i^*=\zeta_i^{(0)}/\nu$ are \cite{BSG20}
\begin{widetext}
\beqa
\label{1.29.2}
\zeta_{i}^*&=&\frac{4\pi^{(d-1)/2}}{d\Gamma\left(\frac{d}{2}\right)}\sum_{j=1}^2
x_j
\left(\frac{\sigma_{ij}}{\sigma_{12}}\right)^{d-1}\mu_{ji}(1+\al_{ij})\theta_i^{-1/2}
\left(1+\theta_{ij}\right)^{1/2}
\left[1-\frac{1}{2}\mu_{ji}(1+\alpha_{ij})(1+\theta_{ij}) \right]\nonumber\\
& &-\frac{4\pi^{d/2}}{d\Gamma\left(\frac{d}{2}\right)}\sum_{j=1}^2 x_j
\left(\frac{\sigma_{ij}}{\sigma_{12}}\right)^{d-1}\mu_{ji}\Delta^*\left[
\frac{2\mu_{ji}\Delta^*}{\sqrt{\pi}}\theta_i^{1/2}\left(1+\theta_{ij}\right)^{1/2}
-1+\mu_{ji}(1+\al_{ij})\left(1+\theta_{ij}\right)\right],
\eeqa
\end{widetext}
where
\beq
\label{1.29.3}
\theta_i=\frac{m_i T}{\overline{m}T_i^{(0)}}, \quad \theta_{ij}=\frac{m_iT_j^{(0)}}{m_jT_i^{(0)}}.
\eeq
The quantity $\theta_{ij}$ gives the ratio between the mean-square velocity of the particles of the component $j$ relative to that of the particles of the component $i$.

\vicente{As mentioned before, the case $\Delta_{ij}\equiv \Delta$ corresponds to the usual case for binary granular mixtures. In the bidisperse case, the unknowns are the steady temperature $T$ (or equivalently, the dimensionless parameter $\Delta^*=\Delta/\sqrt{2T/\overline{m}}$) and the temperature ratio $T_1^{(0)}/T_2^{(0)}$. In the HSS, these two quantities are determined from the conditions
\beq
\label{n1}
\zeta_1^*=0, \quad \zeta_2^*=0,
\eeq
where the explicit forms of the partial cooling rates $\zeta_1^*$ and $\zeta_2^*$ can be easily inferred from Eq.\ \eqref{1.29.2}. The solution to Eqs.\ \eqref{n1} provides the values of $\Delta^*$ (or equivalently, the (reduced) steady temperature $T^*=2T/\overline{m}\Delta^2$) and the temperature ratio $T_1^{(0)}/T_2^{(0)}$ in terms of the parameter space of the system: the mass ratio $m_1/m_2$, the ratio of diameters $\sigma_1/\sigma_2$, the mole fraction $x_1$, and the coefficients of restitution $\al_{11}$, $\al_{22}$, and $\al_{12}$.}
In spite of the Maxwellian approximation  approximation \eqref{1.29.1}, in the low-density regime, the theoretical results obtained for the temperature ratio $T_1^{(0)}/T_2^{(0)}$ and the global temperature $T^*$ compare, in general, quite well with Monte Carlo and molecular dynamics simulations.\cite{BSG20}

It is important to note that the determination of the Navier--Stokes transport coefficients in the HSS requires to evaluate derivatives such as $(\partial \gamma_i/\partial \Delta^*)_s$ and $(\partial \zeta_0^*/\partial \Delta^*)_s$, where $\gamma_i=T_i^{(0)}/T$ and the subscript $s$ means that the derivatives are computed in the steady state (i.e., when $\zeta_1^*=\zeta_2^*=0$). The  explicit forms of the reduced transport coefficients and the cooling rate for the case $\Delta_{11}=\Delta_{22}=\Delta_{12}\equiv \Delta$ in the HSS are displayed in the Appendix \ref{appA}.

\section{Stability analysis}
\label{sec3}

It is quite evident that the knowledge of the complete set of Navier--Stokes transport coefficients of the binary granular mixture opens up the possibility of performing a linear stability analysis of the HSS. This analysis will provide us a critical length $L_c$ beyond which the system becomes unstable. Previous theoretical studies \cite{G05,GMD06,G15} on the HCS in the conventional IHS model have shown that the HCS becomes unstable for long enough wavelength perturbations. These theoretical predictions compare in general quite well with computer simulations. \cite{MDCPH11, MGHEH12,BR13,MGH14} In the case of the $\Delta$-model, a previous work \cite{GBS21} on monocomponent granular gases has shown that the HSS is linearly stable. A natural question arises then as to whether, and if so to what extent, the conclusions achieved in the monocomponent case can be changed in the case of a binary granular mixture.

In the absence of gravity, needless to say the Navier--Stokes hydrodynamic equations \eqref{1.23}--\eqref{1.26} admit the HSS as a solution. This state is a uniform state where $\mathbf{U}_\text{H}=\mathbf{0}$ (without loss of generality), $n_\text{H}\equiv \text{const}$, and $T_\text{H}\equiv \text{const}$. Here, the subscript $H$ means that the hydrodynamic fields are evaluated in the HSS. In addition, as said before, the steady values of the scaled parameter $\Delta^*$ and the temperature ratio $\gamma\equiv T_1^{(0)}/T_2^{(0)}$ are determined from the conditions $\zeta_1^*=\zeta_2^*=0$. The goal of this section is to analyze the stability of the HSS, namely, we want to investigate whether the HSS is linearly stable or not with respect to long enough wavelength perturbations. An answer to this question can be provided when one performs a \emph{linear} stability analysis of the nonlinear Navier--Stokes equations \eqref{1.23}--\eqref{1.26} with respect to the HSS for small initial \emph{inhomogeneous} perturbations.

We assume that the deviations $\delta y_{\beta}({\bf r},t)=y_{\beta}({\bf r},t)-y_{\text{H} \beta}$ are small, where
$\delta y_{\beta}({\bf r},t)$ denotes the deviation of $\{x_1, \mathbf{U},p, T,\}$ from their values in the HSS. For the sake of convenience, we introduce the time and space dimensionless variables:
\beq
\label{2.1}
\tau=\nu_\text{H}t, \quad \boldsymbol{\ell}=\frac{\mathbf{r}}{\ell_\text{H}},
\eeq
where $\nu_\text{H}=n_\text{H} \sigma_{12}^{d-1} \sqrt{2T_\text{H}/\overline{m}}$ and $\ell_\text{H}=1/(n_\text{H}\sigma_{12}^{d-1})$.
The dimensionless time scale $\tau$ is a measure of the average number of collisions per particle
in the time interval between $0$ and $t$. The unit length $\ell_\text{H}=\sqrt{2T_\text{H}/\overline{m}}/\nu_{\text{H}}$ is proportional to the time-independent mean free path of gas particles.

As usual, the linearized hydrodynamic equations for the perturbations $\left\{\delta x_1(\mathbf{r}; t), \delta \mathbf{U}(\mathbf{r}; t), \delta p(\mathbf{r}; t), \delta T(\mathbf{r}; t)\right\}$ are written in the Fourier space. A set of Fourier transformed dimensionless variables are introduced as
\beq
\label{2.2}
\rho_{\mathbf{k}}(\tau)=\frac{\delta x_{1\mathbf{k}}(\tau)}{x_{1\text{H}}}, \quad \mathbf{w}_{\mathbf{k}}(\tau)=\frac{\delta \mathbf{U}_{\mathbf{k}}(\tau)}{v_{\text{th,H}}},
\eeq
\beq
\label{2.3}
\Pi_{\mathbf{k}}(\tau)=\frac{\delta p_{{\bf k}}(\tau)}{p_\text{H}}, \quad
\theta_{{\bf k}}(\tau)=\frac{\delta T_{\mathbf{k}}(\tau)}{T_{\text{H}}},
\eeq
where $v_{\text{th,H}}=\sqrt{2T_\text{H}/\overline{m}}$ and $p_\text{H}=n_\text{H}T_\text{H}$. Here, $\delta y_{\mathbf{k}\beta}\equiv \{\delta n_{\mathbf{k}}(\tau),{\bf w}_{{\bf k}}(\tau), \Pi_\mathbf{k}(\tau), \theta_{{\bf k}}(\tau)\}$ is defined as
\begin{equation}
\label{2.3.1}
\delta y_{\mathbf{k}\beta}(\tau)=\int d{\boldsymbol {\ell}}\;
e^{-\imath \mathbf{k}\cdot {\boldsymbol {\ell}}}\delta y_{\beta}
({\boldsymbol {\ell}},\tau).
\end{equation}
Note that in Eq.\ (\ref{2.3}) the wave vector $\mathbf{k}$ is dimensionless.

In terms of the dimensionless variables \eqref{2.2} and \eqref{2.3.1}, it is quite apparent that the $d-1$ transverse velocity components ${\bf w}_{{\bf k}\perp}={\bf w}_{{\bf k}}-({\bf w}_{{\bf k}}\cdot
\widehat{{\bf k}})\widehat{{\bf k}}$ (orthogonal to the wave vector ${\bf k}$)
decouple from the other four modes and they verify a degenerate $d-1$ differential equations given by
\beq
\label{2.3.2}
\frac{\partial \mathbf{w}_{{\bf k}\perp}}{\partial \tau}+\frac{1+\mu}{4(x_2+\mu x_1)}\eta^* k^2 \mathbf{w}_{{\bf k}\perp}=0,
\eeq
where $\mu=m_1/m_2$ is the mass ratio and $x_i=n_{i}/n$. Since $\eta^*$ does not depend on time in the HSS, the solution to Eq.\ \eqref{2.3.2} is
\beq
\label{2.4}
\mathbf{w}_{{\bf k}\perp}(\tau)=\mathbf{w}_{{\bf k}\perp}(0)\exp\left(-\frac{1+\mu}{4(x_2+\mu x_1)}\eta^* k^2\tau\right).
\eeq
Thus, the $d-1$ transversal shear modes ${\bf w}_{{\bf k}\perp}(\tau)$ are linearly stable because the shear viscosity $\eta^*$ is always positive [see Eq.\ \eqref{a5}].

The set of differential equations for the four longitudinal modes $\rho_\mathbf{k}$, $\theta_\mathbf{k}$, $\Pi_\mathbf{k}$, and $\mathbf{w}_{{\bf k}||}$ (parallel to $\mathbf{k}$) is more intricate. In matrix form, this set can be written as
\begin{equation}
\frac{\partial \delta z_{\mathbf{k}\alpha }(\tau )}{\partial \tau }=\left(
M_{\alpha \beta }^{(0)}+\imath kM_{\alpha \beta }^{(1)}+k^{2}M_{\alpha \beta
}^{(2)}\right) \delta z_{\mathbf{k}\beta }(\tau ),  \label{2.5}
\end{equation}
where now $\delta z_{\mathbf{k}\alpha }(\tau )$ denotes the four variables $
\left( \rho _{\mathbf{k}},\theta _{\mathbf{k}},\Pi _{\mathbf{k}},w_{\mathbf{
k }||}\right) $. The matrices in Eq.\ \eqref{2.5} are
\begin{equation}
M^{(0)}=\left(
\begin{array}{cccc}
0 & 0 & 0 & 0 \\
-A & B& 0&0\\
-A & B& 0&0\\
0 & 0 & 0 &0
\end{array}
\right) ,  \label{2.6}
\end{equation}
\begin{equation}
M^{(1)}=\left(
\begin{array}{cccc}
0 & 0 & 0 & 0 \\
0 & 0 & 0 & -\left(\frac{2}{d}+\zeta_U\right) \\
0 & 0 & 0 & -\left(\frac{d+2}{d}+\zeta_U\right) \\
0 & 0 & -\frac{1}{4}\frac{1+\mu}{x_{1}\mu +x_{2}} & 0
\end{array}
\right) ,  \label{2.7}
\end{equation}
\vspace{0.25cm}
\begin{widetext}
\begin{equation}
M^{(2)}=\left(
\begin{array}{cccc}
-\frac{1}{4}\frac{\mu x_{1}+x_{2}}{\mu_{12}}D^{\ast} & -\frac{1}{4 x_1}\frac{\mu x_{1}+x_{2}}{\mu_{12}}D_T^* & -\frac{1}{4x_1}\frac{\mu x_{1}+x_{2}}{\mu_{12}}D_{p}^{\ast} & 0 \\
x_{1}\left(\frac{1-\mu}{4\mu_{12}}D^*- \frac{2}{d}D^{''*}\right)
& \frac{1-\mu}{4\mu_{12}}D_T^{\ast}-\frac{2}{d}
\lambda^{\ast } & \frac{1-\mu}{4 \mu_{12}}D_p^*-
\frac{2}{d}L^{\ast}& 0 \\
-\frac{2}{d}x_{1}D^{\prime \prime }{}^{\ast} & -\frac{2}{d} \lambda^{\ast}
& -\frac{2}{d}L^{\ast } & 0 \\
0 & 0 & 0 & -\frac{d-1}{2d}\frac{1+\mu}{\mu x_1+x_2}\eta^{\ast}
\end{array}
\right) .  \label{2.8}
\end{equation}
\end{widetext}
In Eq.\ \eqref{2.6}, we have introduced the (dimensionless) quantities
\beq
\label{2.9}
A=x_1\Bigg\{\left(\frac{\partial \zeta_2^*}{\partial x_1}\right)_s+x_1\gamma_1\left[\left(\frac{\partial \zeta_1^*}{\partial x_1}\right)_s-\left(\frac{\partial \zeta_2^*}{\partial x_1}\right)_s\right]\Bigg\},
\eeq
\beq
\label{2.10}
B=\frac{1}{2}\Delta_s^*\Bigg\{\left(\frac{\partial \zeta_2^*}{\partial \Delta^*}\right)_s+x_1\gamma_1\left[\left(\frac{\partial \zeta_1^*}{\partial \Delta^*}\right)_s-\left(\frac{\partial \zeta_2^*}{\partial \Delta^*}\right)_s\right]\Bigg\}.
\eeq
As in the case of the transverse modes, the subscript H has been
suppressed in Eqs. \eqref{2.6}--\eqref{2.10} for the sake of brevity. Moreover, as mentioned before, the derivatives appearing in Eqs.\ \eqref{2.9} and \eqref{2.10} have been evaluated in Ref.\ \onlinecite{GBS21} in the general case. Explicit forms for these derivatives are displayed in the Appendix \ref{appA} in the particular case $\Delta_{11}=\Delta_{22}=\Delta_{12}\equiv \Delta$.

For mechanically equivalent particles ($m_1=m_2$, $\sigma_1=\sigma_2$, and $\al_{ij}\equiv \al$), $D_p^*=D_T^*=0$, which implies $L^*=\lambda^*=0$ in the first Sonine approximation. Moreover, in this limiting case,  $A=0$, $B=(\Delta^*/2)(\partial \zeta_0^*/\partial \Delta^*)_s$, and the results are consistent with those obtained for monocomponent granular gases. \cite{GBS21a} Here,
\beq
\label{n2}
\zeta_0^*=\frac{\sqrt{2}\pi^{\frac{d-1}{2}}}{d\Gamma\left(\frac{d}{2}\right)}  \left(1-\al^2-2\Delta^{*2}-\sqrt{2\pi}\al \Delta^*\right)
\eeq
is the reduced cooling rate for a monocomponent granular gas.

\begin{figure}[h!]
\centering
\includegraphics[width=0.4\textwidth]{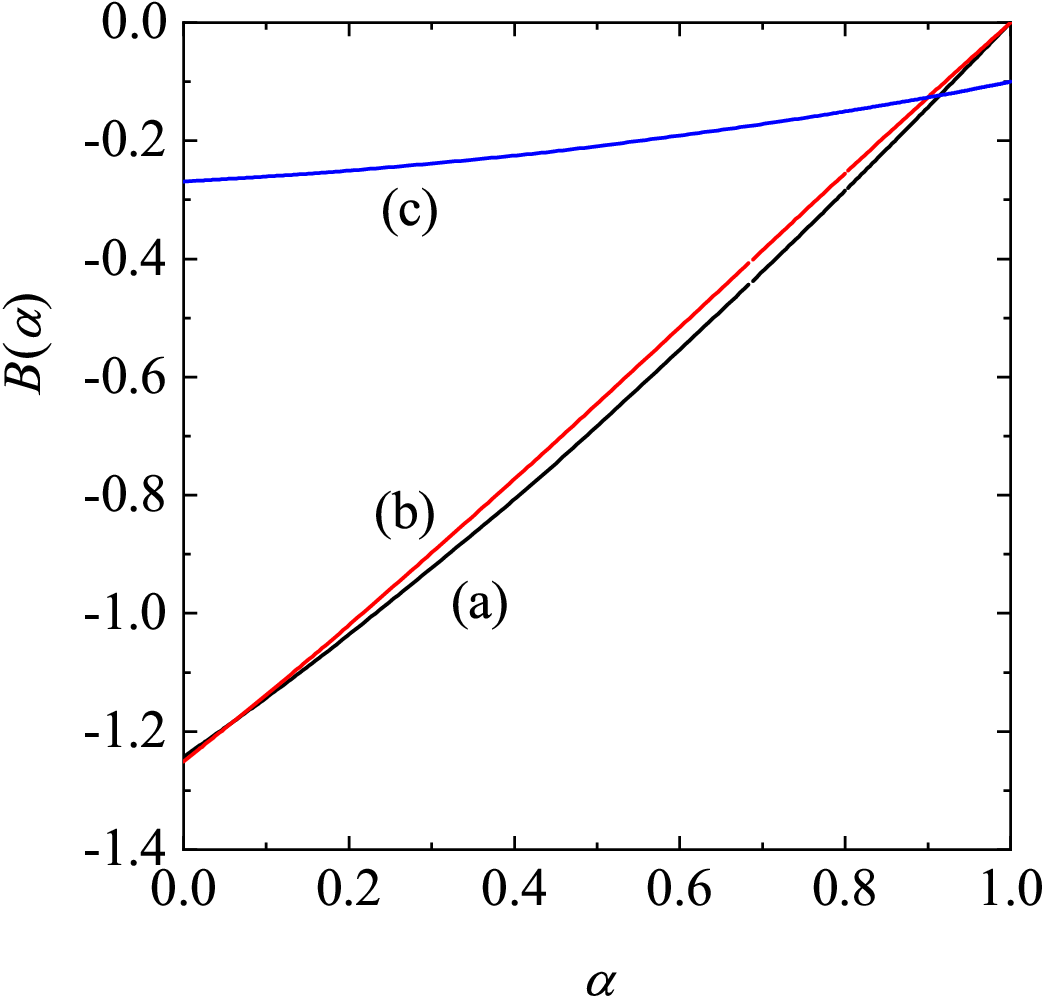}
\caption{Dependence of the eigenvalue $B$ on the coefficient of restitution $\al$ for a two-dimensional system and three different granular binary mixtures: $x_1=0.5$, $\sigma_1/\sigma_2=2$, $m_1/m_2=4$, and $\al_{ij}\equiv \al$ (a); $x_1=0.5$, $\sigma_1/\sigma_2=0.2$, $m_1/m_2=0.5$, and $\al_{ij}\equiv \al$ (b); and $x_1=0.2$, $\sigma_1/\sigma_2=m_1/m_2=1$, $\al_{22}$ =0.9 , $\al_{11}\equiv \al$, and $\al_{12}=(\al_{22}+\al)/2$. (c)}
\label{fig1}
\end{figure}

The time evolution of the fourth longitudinal modes has the form $e^{s_n(k) \tau}$ for $n=$1, 2, 3, and 4. The quantities $s_n(k)$ are the eigenvalues of the matrix
\beq
\label{2.10.0}
M_{\al \beta}=M_{\al \beta}^{(0)}+\imath k M_{\al \beta}^{(1)}+k^2 M_{\al \beta}^{(2)}.
\eeq
Thus, the eigenvalues $s_n(k)$ are the solutions of the quartic equation
\beq
\label{2.11}
\det \left(\mathsf{M}-s \openone \right)=0,
\eeq
where $\openone$ is the matrix identity.

It is quite apparent that the determination of the dependence of the eigenvalues $s_n(k)$ on the (dimensionless) wave vector $k$ and the parameters of the mixture is not really a simple problem. Therefore, to gain some insight into the general problem, it is convenient to study first the solution to the quartic equation \eqref{2.11}  \vicente{in the extreme long wavelength limit, $k=0$}.

\begin{figure}[h!]
\centering
\includegraphics[width=0.4\textwidth]{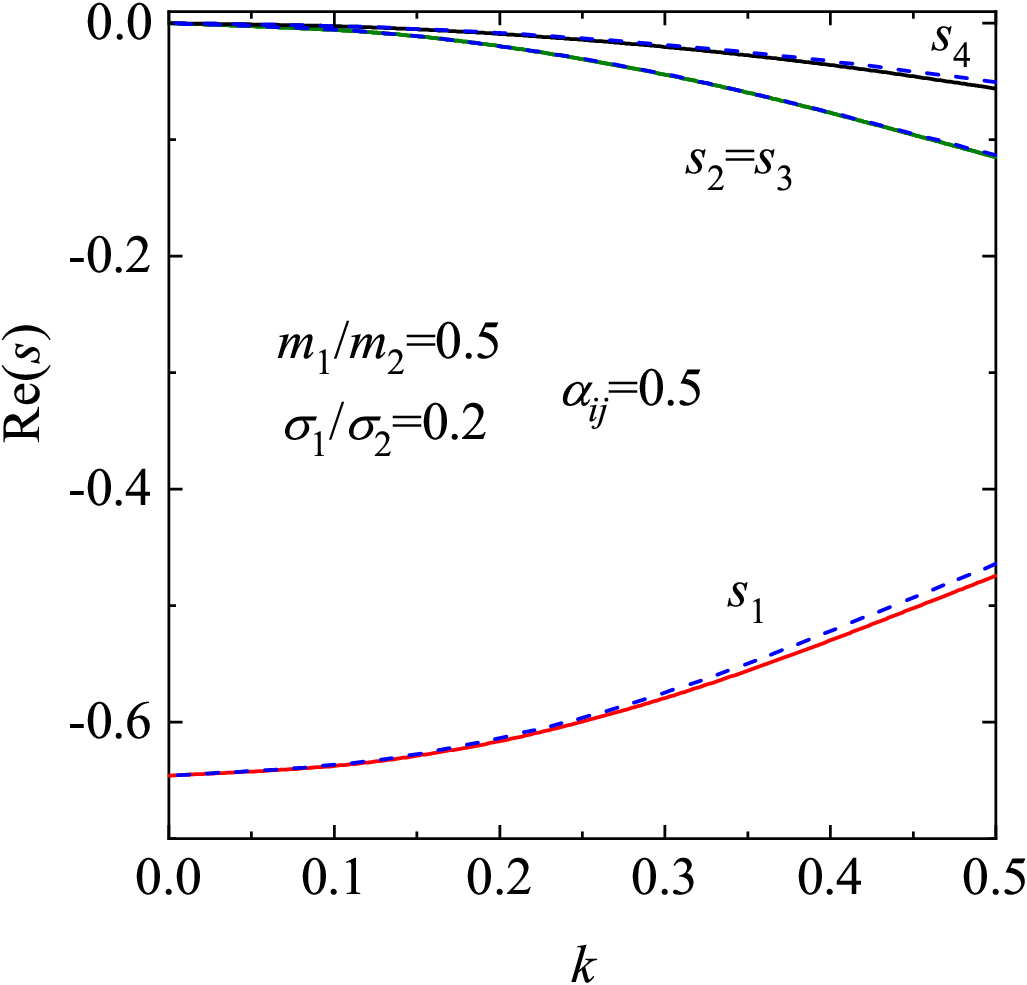}
\caption{Real parts of the longitudinal eigenvalues $s_i$ as functions of the wave number $k$ for a two-dimensional granular binary mixture with $x_1=0.5$, $\sigma_1/\sigma_2=0.2$, $m_1/m_2=0.5$ and the (common) coefficient of restitution $\al_{ij}\equiv 0.5$. \vicente{The dashed lines correspond to the results obtained for monodisperse granular gases}.}
\label{fig2}
\end{figure}

\subsection{\vicente{Extreme long wavelength limit ($k=0$)}}

When $k=0$, the square matrix $\mathsf{M}$ reduces to $\mathsf{M}^{(0)}$ whose eigenvalues are   \beq
\label{2.12}
s_{||}=\left\{0,0,0,B\right\}.
\eeq
According to Eq.\ \eqref{2.10}, the dependence of $B$ on the parameter space of the system is in general complex. A simple situation corresponds to the case of mechanically equivalent particles where $\zeta_1^*=\zeta_2^*=\zeta_0^*$ and so, \cite{GBS18}
\beq
\label{2.13}
\left(\frac{\partial \zeta_0^*}{\partial \Delta^*}\right)_s=-\vicenteg{\frac{\sqrt{2}\pi^{\frac{d-1}{2}}}{d\Gamma\left(\frac{d}{2}\right)}} \left(\sqrt{2\pi}\al+4\Delta^*\right).
\eeq
Thus,
\beq
\label{2.14}
B=-\vicenteg{\frac{\pi^{\frac{d-1}{2}}}{\sqrt{2}d\Gamma\left(\frac{d}{2}\right)}} \Delta^*\left(\sqrt{2\pi}\al+4\Delta^*\right)<0,
\eeq
and the longitudinal modes are linearly stable in agreement with previous results. \cite{GBS21a}

In the case of granular mixtures, a detailed study of the dependence of the quantity $B$ on the parameters of the mixture shows that $B$ is always negative. As a consequence, all the longitudinal modes are stable \vicente{when $k=0$} in the $\Delta$-model \vicente{for the choice $\Delta_{ij}=\Delta$}. As an illustration, Fig.\ \ref{fig1} shows the dependence of $B$ on the coefficient of restitution $\al_{11}\equiv \al$ for three different mixtures. We clearly observe that the eigenvalue $B$ is always negative; its magnitude increases with decreasing $\al$.

%\subsection{Small values of the wave number}

\subsection{General case}

The study at finite wave vectors is
quite complex and requires to numerically solve Eq.\ \eqref{2.11}. This
is a quite hard task due to the large number of parameters involved in the system. On the other hand, analytical forms of the eigenvalues of the matrix $\mathsf{M}$ can be provided when one considers the limit of small wave numbers ($k\to 0$). This study is relevant because the Navier--Stokes hydrodynamic equations apply to second order in $k$. In the limit $k\to 0$, the solution to Eq.\ \eqref{2.11} can be written as
\beq
\label{2.14.1}
s_n(k)=s_n^{(0)}+k s_n^{(1)}+k^2 s_n^{(2)}+\cdots.
\eeq
The coefficients $s_n^{(0)}$, $s_n^{(1)}$, and $s_n^{(2)}$ can be obtained by substituting the expansion \eqref{2.14.1} into the quartic equation \eqref{2.11}. Their explicit expressions are displayed in the Appendix \ref{appB}.

As said before, beyond the limit $k\to 0$, the eigenvalues of $\mathsf{M}$ must be numerically determined.
A careful study of the dependence of the eigenvalues of the matrix $\mathsf{M}$ on the parameters of the mixture shows that the real part of \emph{all} the eigenvalues is \emph{negative} and hence, the HSS is linearly stable in the complete range of values of the wave number $k$ studied. This result contrasts with the one obtained in the conventional IHS model \cite{GMD06} but agrees with previous works \cite{BBGM16,GBS21} carried out in the context of the $\Delta$-model for monocomponent granular gases.  As an illustration, Fig.\ \ref{fig2} shows the real parts of the eigenvalues $s_i$ ($i=1, 2, 3, 4$) as functions of the (dimensionless) wave number $k$ for a two-dimensional granular binary mixture with a concentration $x_1=0.5$, a diameter ratio $\sigma_1/\sigma_2=0.2$, and a (common) coefficient of restitution $\al_{ij}=0.5$. \vicente{The corresponding results for monodisperse granular gases are also plotted for the sake of comparison.} We observe that two of the modes (denoted as $s_2$ and $s_3$) are a complex conjugate pair of propagating modes [$\text{Re}(s_2)=\text{Re}(s_3)$] while the other two modes ($s_1$ and $s_4$) are real for all values of the wave number. \vicente{This feature is also present in the monocomponent case.} We see that the magnitude of $s_4$ is very small while the mode $s_1$ ($s_2$) increases (decreases) with increasing the wave number. \vicente{Figure \ref{fig2} also highlights that the impact of the disparity in masses and/or diameters does not play a significant role on the dependence of the eigenvalues $s_i$ on the wave number $k$ since the results for monocomponent granular gases are quite close to the ones found for bidisperse systems.}

\section{Thermal diffusion segregation}
\label{sec4}

Another interesting application of the results derived in Ref.\ \onlinecite{GBS21} is the study of segregation induced by both a thermal gradient and gravity. Segregation and mixing of dissimilar grains is one of the most interesting problems in granular mixtures, not only from a fundamental point of view but also from a more practical perspective. This problem has spawned a number of important experimental, computational, and theoretical works in the field of granular media, especially when the system is fluidized by vibrating walls.

\begin{figure}
\includegraphics[width=0.8\columnwidth,angle=0]{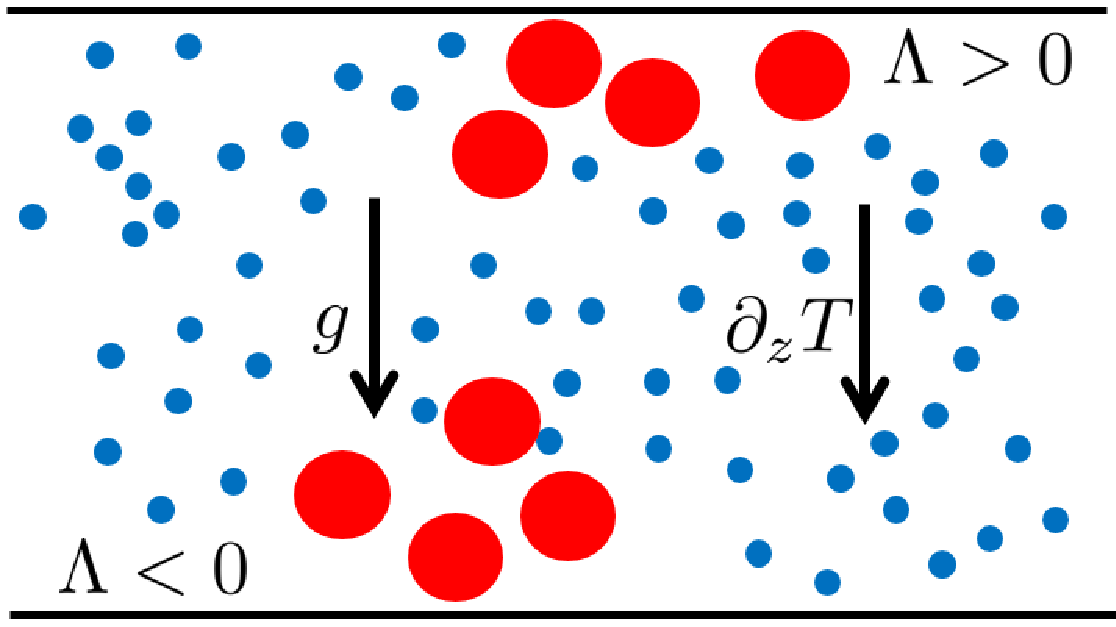}
\caption{(color online) A sketch representation of the segregation problem studied here. Blue circles represent the smaller components of the mixture while the red circles correspond to the larger ones. When the thermal diffusion factor is positive (negative), the smaller particles tend to accumulate near the cold (hotter) wall.
\label{sketch}}
\end{figure}

\vicente{Thermal diffusion is the phenomenon caused by the relative motion of the components of a mixture due to the presence of a thermal gradient.
Due to the motion of the species of the mixture, concentration gradients appear in the mixture. These gradients give rise to diffusion processes.
A steady state is reached where the segregation effect arising from thermal diffusion is compensated for the mixing effect of ordinary diffusion.} \cite{KCL87} The amount of segregation parallel to the thermal gradient may be characterized by the thermal diffusion factor $\Lambda$. \vicente{This quantity (which has been widely studied in the case of molecular mixtures) provides a convenient measure of the separation between the components of a mixture. The thermal diffusion factor} is defined in an inhomogeneous non-convecting (${\bf U}={\bf 0}$) steady state with zero mass flux (${\bf j}_1={\bf 0}$) through the relation
\begin{equation}
\label{3.1}
-\Lambda\frac{\partial \ln T}{\partial z} =\frac{\partial}{\partial z}\ln
\left(\frac{n_1}{n_2}\right),
\end{equation}
where gradients only along the $z$ axis (vertical direction) have been assumed for
simplicity. \vicente{In addition, without loss of generality, we assume that $\sigma_1>\sigma_2$
and also that gravity and the thermal gradient point in parallel directions (i.e., the bottom plate is hotter than the top plate, $\partial_z T < 0$, see the sketch of Fig.\ \ref{sketch})}.

In the above geometry, according to Eq.\ (\ref{3.1}), when $\Lambda >0$ the  larger particles $1$ tend to rise with respect to the
smaller particles $2$ (i.e., $\partial_z\ln (n_1/n_2)>0$). On the other hand, when $\Lambda <0$, the
larger particles fall with respect to the smaller particles (i.e., $\partial_z\ln
(n_1/n_2)<0$). In summary, when $\Lambda> 0$, the larger particles accumulate at the top of the sample
(cold plate), while if $\Lambda<0$, the larger particles accumulate at the bottom of the sample
(hot plate).

We write now the factor $\Lambda$ in terms of the (dimensionless) diffusion coefficients ($D^*$, $D_p^*$, and $D_T^*$) and the gravity field. Since $\mathbf{U}=\mathbf{0}$, then the momentum balance equation \eqref{1.11} leads to
\beq
\label{3.2}
\partial_z p=-\rho g.
\eeq
Moreover, the condition $j_{1,z}=0$ yields
\beq
\label{3.3}
\partial_z x_1=-\frac{D_p^*}{D^*}\partial_z \ln p-\frac{D_T^*}{D^*}\partial_z \ln T,
\eeq
where the dimensionless diffusion coefficients are defined by Eq.\ \eqref{1.27}. Substitution of the relations \eqref{3.2} and \eqref{3.3} into Eq.\ \eqref{3.1} gives the expression
\beq
\label{3.4}
\Lambda=\frac{1}{x_1x_2}\frac{D_T^*-g^* D_p^*}{D^*},
\eeq
where $g^*=\rho g/n \partial_z T<0$ is a dimensionless parameter measuring the gravity relative to the temperature gradient.
The condition $\Lambda=0$ provides the criterion for the transition from $\Lambda>0$ to $\Lambda<0$. Since the (scaled) diffusion coefficient $D^*$ is positive [see Eq.\ \eqref{a3}], then the sign of $\Lambda$ is the same as that of the difference $D_T^*-g^* D_p^*$. In particular, in the absence of gravity ($g^*=0$), the sign of $\Lambda$ coincides with the sign of the (scaled) thermal diffusion coefficient $D_T^*$. This segregation criterion differs from the one obtained in previous works for free cooling granular gases \cite{G11} or when the gas is driven by a stochastic thermostat.\cite{G06,G08a,G09,VGK14}

According to Eq.\ \eqref{3.4}, the segregation is driven and sustained by both gravity and temperature gradients. Although in our work we have assumed that gravity and thermal gradient are of the same order of magnitude, it is interesting for illustrative purposes to separate the influence of each one of the terms appearing in Eq.\ \eqref{3.4} on segregation. Thus, the specific cases of absence of gravity ($g=0$ but $\partial_z T\neq 0$) or thermalized systems ($\partial_zT\to 0$ but $g\neq 0$) will be considered in Secs.\ \ref{IVA} and \ref{IVB}.

\begin{figure}[h!]
\centering
\includegraphics[width=0.4\textwidth]{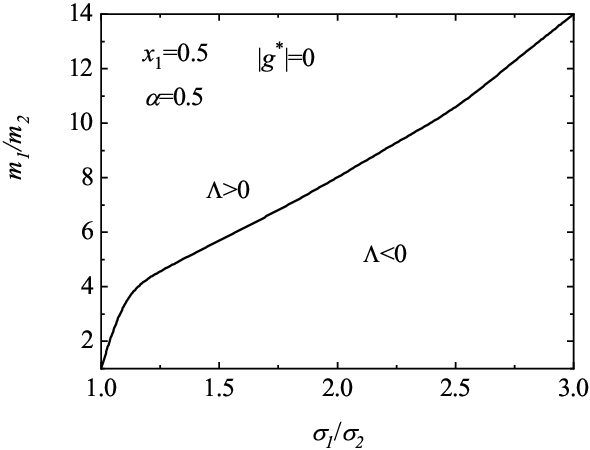}
\caption{\vicenteg{Phase diagram for the marginal segregation curve ($\Lambda=0$) in the $\left(\sigma_1/\sigma_2, m_1/m_2\right)$-plane for a two-dimensional system with $x_1=0.5$ and a (common) coefficient of restitution $\al=0.5$. The limiting case  $|g^*|=0$ is considered}.}
\label{figg0}
\end{figure}

\subsection{Absence of gravity ($|g^*|=0$)}
\label{IVA}

We study first the segregation of two species of grains in the presence of a temperature gradient, but in the absence
of gravity ($g=0$). In this limiting case ($|g^*|=0$), $\text{sign}(\Lambda)=\text{sign}(D_T^*)$, where $D_T^*$ is given by Eq.\ \eqref{a2}. An exhaustive analysis on the dependence of the transport coefficient $D_T^*$ on the parameter space of the system shows that $D_T^*$ \vicenteg{may be positive or negative. Thus, depending on the parameter space of the system, partial separation is observed where the larger components move towards the cold or the hot wall} (see Fig.\ \ref{sketch}). \vicenteg{As an illustration, Fig.\ \ref{figg0} shows the marginal segregation curve ($\Lambda=0$) delineating the regimes between $\Lambda>0$ and $\Lambda<0$ in the $\left(\sigma_1/\sigma_2, m_1/m_2\right)$-plane for a two-dimensional system with $x_1=0.5$ and a (common) coefficient of restitution $\al_{ij}\equiv \al=0.5$. We observe that the region where the thermal diffusion factor becomes negative (larger particles segregate towards the hot/bottom plate) is dominant at both small mass ratios $m_1/m_2$ and/or large diameter ratios $\sigma_1/\sigma_2$.}  It must be remarked that the results derived here in the $\Delta$-model contrast with those obtained when the granular mixture is driven with a stochastic thermostat (see Fig.\ 5 of Ref.\ \onlinecite{VGK14}).

\subsection{Thermalized systems ($\partial_z T\to 0$)}
\label{IVB}

\begin{figure}[h!]
\centering
\includegraphics[width=0.4\textwidth]{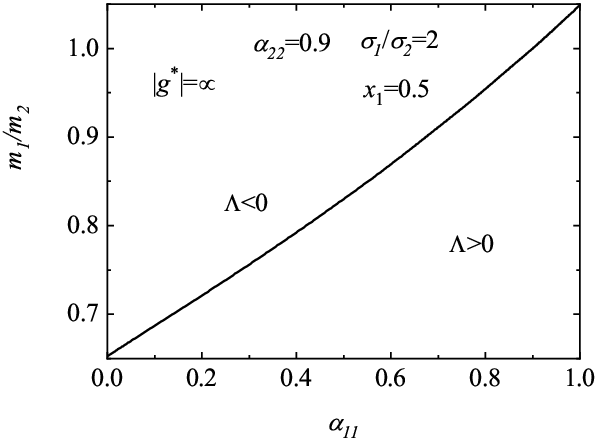}
\caption{Plot of the dependence of the marginal segregation curve ($\Lambda=0$) on the coefficient of restitution $\al_{11}$ for a two-dimensional system with $x_1=0.5$, $\sigma_1/\sigma_2=2$, $\al_{22}=0.9$, and $\al_{12}=(\al_{11}+\al_{12})/2$. The limiting case  $|g^*|\to \infty$ is considered.}
\label{fig3}
\end{figure}

We consider now a situation where the inhomogeneities in the temperature are neglected but gravity is different from zero.
In this limiting case, the segregation is only driven by the gravitational force. This situation (gravity dominates the temperature gradient) can be achieved in the shaken or sheared systems employed in
numerical simulations and physical experiments. \cite{HQL01,BEKR03,SBKR05} Under these conditions ($|g^*|\to \infty$), $\Lambda\approx D_p^* |g^*|$ and so the sign of $\Lambda$ is the same as that of the pressure diffusion coefficient $D_p^*$. Figure \ref{fig3} shows the dependence of the marginal segregation curve ($\Lambda=0$) on $\al_{11}$ for a two-dimensional system with $x_1=0.5$, $\sigma_1/\sigma_2=2$, $\al_{22}=0.9$, and $\al_{12}=(\al_{11}+\al_{12})/2$. For a given value of the coefficient of restitution $\al_{11}$, it is quite apparent that $\Lambda$ is always negative (larger particles accumulate near the hot plate) when the larger particles are heavier than that of the smaller ones ($m_1>m_2$). However, when $m_1<m_2$, we see that the size of the region where $\Lambda$ becomes positive increases with decreasing the mass ratio $m_1/m_2$.

\subsection{General case}

\begin{figure}[h!]
\centering
\includegraphics[width=0.4\textwidth]{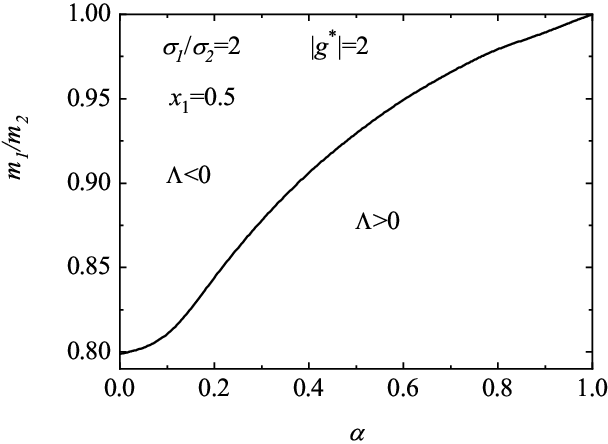}
\caption{Plot of the dependence of the marginal segregation curve ($\Lambda=0$) on the (common) coefficient of restitution $\al_{ij}\equiv \al$ for a two-dimensional system with $x_1=0.5$, $\sigma_1/\sigma_2=2$, and $|g^*|=2$.}
\label{fig4}
\end{figure}

Finally, we consider the general case for finite values of the reduced gravity. Figure \ref{fig4} illustrates this situation; we plot the marginal segregation curve ($\Lambda=0$) versus the (common) coefficient of restitution $\al$ for a two-dimensional system with $x_1=0.5$, $\sigma_1/\sigma_2=2$, and $|g^*|=2$. In contrast to Fig.\ \ref{fig3}, we observe that the region $\Lambda>0$ appears essentially for both \vicenteg{small} mass ratio $m_1/m_2$ and/or \vicenteg{moderate} inelasticity.

\vicente{We are not aware of available computer simulations in the context of the $\Delta$-model to compare the theoretical predictions displayed in the marginal curves of Figs.\ \ref{fig3} and \ref{fig4} against simulation data. We expect that the present results stimulate the performance of such simulations to assess the degree of reliability of the theoretical results derived here for thermal diffusion segregation.}

\section{Summary and discussion}
\label{sec5}

In this paper we report the stability analysis of the hydrodynamic equations and the study of  thermal segregation  for a mixture of granular particles evolving under the so-called $\Delta$-model. This model is an extension of the usual IHS model where in every particle collision, we add an amount $\Delta$ to the normal velocities of the colliding particles. This mechanism injects energy into the system, mimicking the dynamics of vibrofluidized quasi-two dimensional setups and avoiding the problems of other thermostats. With both an energy injection and collisional dissipation (characterized by the normal restitution coefficients, $\alpha_{ij}$), the systems reaches a homogeneous  steady state (HSS).

The first goal of the present paper has been to perform a linear stability analysis of such HSS.
The starting point is the set of hydrodynamic equations for a mixture of two types of inelastic particles, characterized by their masses ($m_1, m_2$), diameters ($\sigma_1, \sigma_2$) and three restitution coefficients ($\alpha_{11}$ and $\alpha_{22}$ for collisions between 1--1 and 2--2 particles, and $\alpha_{12}$ for collisions between 1--2 particles). Gravity  $g$ is also included in the derivation of the hydrodynamic equations; this quantity appears in the balance equation for the flow velocity $\mathbf{U}$. The hydrodynamic equations derived from the Boltzmann equation form a set of $d+3$ differential equations for the partial number densities $n_1$ and $n_2$, the flow velocity $\mathbf{U}$, and the granular temperature $T$. The temperature $T$ is not strictly a conserved quantity, but its equation contains a term that balances collisional dissipation (accounted by the set of the restitution coefficients $\alpha_{ij}$) with energy injection (characterized by the $\Delta$ coefficient). Despite this set of variables is correct, it is, however, more convenient to express the hydrodynamic equations in terms of the mole fraction $x_1=n_1/(n_1+n_2)$, the mean flow velocity $\mathbf U$, the pressure $p=n T$ and the temperature $T$. In addition,
the temperatures of each species are in general different, \cite{BSG20} finding a nonequipartion of energy which is usual in granular mixtures.
The hydrodynamic equations are closed with the constitutive relations to linear order in the gradients, described in Eqs. \eqref{1.19}--\eqref{1.22}, where the transport coefficients have been explicitly  calculated in Ref. \onlinecite{GBS21}.

Then, in section III, we proceed to carry out the stability analysis of the HSS in the absence of gravity for small deviations of the fields with respect to the HSS solution. The HSS state is  characterized by a homogeneous system at rest (${\mathbf U}=0$). Deviations are written in Fourier space and relaxation rates are expanded in series of the wave vector. We find that the $d-1$ transverse velocity modes are stable. Moreover, we provide explicit expressions for the relaxation rates (up to second order in the wave vector $k$) for the 4 longitudinal modes. The analytical expressions are rather involved, but numerical analysis for finite $k$ shows that the real parts of them are negative for all values of the system parameters, indicating that the HSS is linearly stable for any long wavelength excitation.  Such finding coincides with earlier stability analyses for monocomponent granular gases using the same model, \cite{GBS21a} but disagrees with the IHS (purely dissipative models) where the HCS is unstable for sufficiently long wavelengths.

\vicente{It is quite apparent that the results obtained for the stability of the HSS has been derived in the particular case where $\Delta_{11}=\Delta_{22}=\Delta_{12}$. As said in Sec.\ \ref{sec2}, this case is quite interesting since it refers to a binary mixture where their components differ in masses, diameters, and coefficients of restitution and not in their energy injection at collisions. The extension of these results when $\Delta_{11}\neq\Delta_{22}\neq \Delta_{12}$ is quite cumbersome since it requires to evaluate the Navier--Stokes transport coefficients of the mixture when $\partial \gamma_i/\partial \Delta_{11}^*\neq \partial \gamma_i/\partial \Delta_{22}^*\neq \partial \gamma_i/\partial \Delta_{12}^*$ and $\partial \zeta_0^*/\partial \Delta_{11}^*\neq \partial \zeta_0^*/\partial \Delta_{22}^*\neq \partial \zeta_0^*/\partial \Delta_{12}^*$. On the other hand, our main expectation is that the conclusion reached here for the stability of the HSS is quite robust and independent of the choice of $\Delta_{ij}$.}

\vicente{Apart from studying the stability of the HSS, we have also analyzed} thermal segregation induced by the competition of gravity and a thermal gradient. Segregation is unveiled by the thermal diffusion factor, $\Lambda$, defined in Eq.~(\ref{3.1}). An explicit expression is given in Eq.~(\ref{3.4}). Assuming a hot base of the container and a cold lid, when $\Lambda>0$ larger particles rise to the top of the container (against gravity), while small ones sink to the bottom of the container. When $\Lambda<0$ the effect is the opposite. The value of $\Lambda=0$  represents the locus line of the separation of the two opposite behaviors.

The limiting cases of absence of gravity ($g\to 0$) and no thermal gradient ($\partial_z T\to0$) are then analyzed. \vicenteg{In both limiting cases,} the behavior is rich, finding a region of $\Lambda>0$ and another of $\Lambda<0$, depending on collisional dissipation and mass and/or diameter ratios. Finally we consider the case of gravity and thermal gradient, when a much complex dependence on the parameters appears.

\vicente{One of the main limitations of the present study is the absence of computer simulations to assess the degree of accuracy of the theoretical results derived for the stability of the HSS and the thermal diffusion segregation problem. In particular, the assessment of the thermal diffusion factor $\Lambda$ for the $\Delta$-model is still an open interesting issue.
However, its determination in computer simulations in the Navier--Stokes domain may be a difficult problem due essentially to the inherent coupling present in steady states for granular gases between spatial gradients and collisional cooling. \cite{SGD04} In any case, previous DSMC simulations \cite{VGK14} for driven granular mixtures have shown a good agreement between the results derived for $\Lambda$ from kinetic theory and numerical simulations, even for strong inelasticity. We expect that this good agreement is also present in the case of the $\Delta$-model. We plan to implement a numerical code in the near future to measure the thermal diffusion factor in the Navier--Stokes regime.}

In conclusion, the present paper extends the understanding of granular matter in quasi-two-dimensional mixtures of granular particles.  We obtain that the HSS created by the injection of energy is linearly stable. This stability result contrasts with the behavior of the HCS in the conventional IHS model, showing that energy injection due to $\Delta$, tends to break velocity correlations and then  stabilizes the system. Finally, we study the segregation due to both a thermal gradient and gravity, deriving a segregation criterion. The predictions could be tested against experiments of  vibrated granular mixtures and test if the $\Delta$-model can mimic real systems where the energy injection is carried out through the vibrating plates. The effect of gravity can be included for example, by slightly tilting the setup and the temperature gradient can be induced by modulating the surface roughness of the vibrating plates.

\section*{Acknowledgments}
The work of V.G. is supported from Grant No. PID2020-112936GB-I00 funded by
MCIN/AEI/10.13039/501100011033 and from Grant No.
IB20079 funded by Junta de Extremadura (Spain) and by
ERDF “A Way of Making Europe.” The work of R.B. is supported from Grant Number PID2020-113455GB-I00.
R.S. is supported by the Fondecyt Grant
No. 1220536, from ANID, Chile.

\vspace{0.5cm}

\textbf{AUTHOR DECLARATIONS}
\textbf{Conflict of Interest}

The authors have no conflicts to disclose.

\vspace{0.5cm}

\textbf{Author Contributions}\\

\textbf{Vicente Garz\'o}: Formal analysis (equal); Investigation
(equal); Software (equal); Writing – review and editing (equal).\textbf{Ricardo Brito}: Conceptualization (equal); Investigation (equal); Writing – original draft (equal). \textbf{Rodrigo Soto}: Conceptualization (equal); Investigation (equal); Writing – original draft (equal).

\vspace{0.5cm}

\textbf{DATA AVAILABILITY}

The data that support the findings of this study are available from the corresponding author upon reasonable request.

\appendix
\section{Expressions of the transport coefficients and the cooling rate in the HSS}
\label{appA}

In this Appendix, we provide the explicit expressions of the (reduced) transport coefficients $D^*$, $D_p^*$, $D_T^*$, $\eta^*$, $D^{''*}$, $L^*$, $\lambda^*$, and $\zeta_U$ in the HSS. The transport coefficients associated with the mass flux are
\beq
\label{a1}
D_p^*=\frac{x_1}{\nu_D^*}\Bigg(\gamma_1-\frac{\mu}{x_2+\mu x_1}\Bigg),
\eeq
\beq
\label{a2}
D_T^*=-\frac{x_1 \Delta^*\left(\frac{\partial \gamma_1}{\partial \Delta^*}\right)+\Delta^*\left(\frac{\partial \zeta_0^*}{\partial \Delta^*}\right) D_p^*}{2\nu_D^*+\Delta^*\left(\frac{\partial \zeta_0^*}{\partial \Delta^*}\right)},
\eeq
\beq
\label{a3}
D^*=\frac{\gamma_1+x_1\left(\frac{\partial \gamma_1}{\partial x_1}\right)+\left(D_p^*+D_T^*\right)\left(\frac{\partial \zeta_0^*}{\partial x_1}\right)}{\nu_D^*},
\eeq
where $\zeta_0^*=\zeta_{2}^*+x_1 \gamma_{1}\left(\zeta_{1}^*-\zeta_{2}^*\right)$ and
\beqa
\label{a4}
\nu_D^*&=&\frac{2\pi^{(d-1)/2}}{d\Gamma\left(\frac{d}{2}\right)}\left(x_1\mu_{12}+x_2\mu_{21}\right)
\nonumber\\
& & \times
\left[\left(\frac{\theta_1+\theta_2}{\theta_1\theta_2}\right)^{1/2}(1+\al_{12})+\sqrt{\pi}\Delta^*\right],
\eeqa
Here, we recall that $\mu=m_1/m_2$ is the mass ratio and $\theta_i=m_i/(\overline{m}\gamma_{i})$. Henceforth, it is understood that all the quantities displayed in this Appendix are evaluated in the steady state.

The (reduced) shear viscosity $\eta^*$ is given by
\beq
\label{a5}
\eta^*=\frac{\left(\tau_{22}^*-\tau_{21}^*\right)x_1\gamma_1+\left(\tau_{11}^*-\tau_{12}^*\right)x_2\gamma_2}
{\tau_{11}^*\tau_{22}^*-\tau_{12}^*\tau_{21}^*},
\eeq
where the expressions of the (dimensionless) quantities $\tau_{ij}^*=\tau_{ij}/\nu$ can be easily identified from Eqs.\ (C7)--(C10) of the Appendix C of Ref.\ \onlinecite{GBS21}.

In the first Sonine approximation, the expressions of the (reduced) transport coefficients associated with the heat flux are
\beq
\label{a6}
\left\{D^{''*}, L^*, \lambda^*\right\}=\frac{d+2}{4}\left(\frac{\gamma_1}{\mu_{12}}-\frac{\gamma_2}{\mu_{21}}\right)\left\{D^{*}, D_p^*, D_T^*\right\}.
\eeq

The quantity $\zeta_U$ can be written as
\beq
\label{a7}
\zeta_U=\sum_{i=1}^2\xi_i^*\varpi_i^*,
\eeq
where
%\begin{widetext}
\beqa
\label{a8}
\xi_i^{*}&=&\frac{3\pi^{(d-1)/2}}{d\Gamma\left(\frac{d}{2}\right)}\frac{m_i}{\overline{m}\gamma_i}\sum_{j=1}^2 x_i x_j \left(\frac{\sigma_{ij}}{\sigma_{12}}\right)^{d-1}\nonumber\\
& & \times
\mu_{ji}(1-\al_{ij}^2)\left(\theta_i+\theta_j\right)^{1/2}\theta_i^{-3/2}\theta_j^{-1/2}\nonumber\\
& & -\frac{4\pi^{(d-1)/2}}{d\Gamma\left(\frac{d}{2}\right)}x_i \Delta^*\sum_{j=1}^2 x_j \left(\frac{\sigma_{ij}}{\sigma_{12}}\right)^{d-1}
\mu_{ji}\nonumber\\
& & \times \Bigg\{\sqrt{\pi}\al_{ij}+\left(\theta_i+\theta_j\right)^{-1/2}\theta_i^{3/2}\theta_j^{-1/2}\Delta^*\nonumber\\
& & \times
\Big[d-d\left(\theta_i+\theta_j\right)\theta_i^{-1}+(d+1)\theta_i \theta_j^{-1}\Big]\Bigg\},\nonumber\\
\eeqa
\beq
\label{a9}
\varpi_1^*=\frac{1}{d}\frac{\Delta^*\left(\frac{\partial \gamma_1}
{\partial \Delta^*}\right)}{\Lambda_1^*}, \quad \varpi_2^*=-\frac{x_1}{x_2}\varpi_1^*,
\eeq
\beq
\label{a10}
\Lambda_1^*=\omega_{11}^*-\frac{x_1}{x_2} \omega_{12}^*-\Bigg[\frac{1}{2}\Delta^*
\left(\frac{\partial \gamma_1}{\partial \Delta^*}\right)-\gamma_1\Bigg]\Bigg(\xi_1^*-\frac{x_1}{x_2} \xi_2^*\Bigg).
\eeq
In Eq.\ \eqref{a10}, the expressions of $\omega_{11}^*=\omega_{11}/\nu$ and $\omega_{12}^*=\omega_{12}/\nu$ can be easily obtained from Eqs.\ (C12) and (C13), respectively, of the Appendix C of Ref.\ \onlinecite{GBS21}.

According to Eqs.\ \eqref{a1}--\eqref{a3} and \eqref{a10}, it is quite apparent that the diffusion transport coefficients and the first-order contribution to the cooling rate are given in terms of the derivatives $(\partial \gamma_1/\partial \Delta^*)$ and $(\partial \gamma_1/\partial x_1)$. The derivative $(\partial \gamma_1/\partial \Delta^*)$ is  \cite{GBS21,note}
\beq
\label{a11}
\left(\frac{\partial \gamma_1}{\partial \Delta^*}\right)=\frac{\sqrt{Y^2-4X Z}-Y}{2X},
\eeq
where $X=N \Delta^*$,
\beq
\label{a11.1}
Y=M \Delta^*-2 N \gamma_{1}+\gamma_{1}
\left(\frac{\partial \zeta_1^*}{\partial \gamma_1}\right), \quad Z=\gamma_{1}\left(\frac{\partial \zeta_1^*}{\partial \vicenteg{\Delta^*}}\right)-2 M \gamma_{1}.
\eeq
%\beq
%\label{a11.2}
%Z=\gamma_{1}\left(\frac{\partial \zeta_1^*}{\partial \gamma_1}\right)-2 A \gamma_{1}.
%\eeq
Here,
\beq
\label{A}
M=\frac{1}{2}\Bigg[x_1 \gamma_1 \Big(\frac{\partial \zeta_1^*}{\partial \Delta^*}\Big)_{\gamma_1}
+x_2 \gamma_2 \Big(\frac{\partial \zeta_2^*}{\partial \Delta^*}\Big)_{\gamma_1}\Bigg],
\eeq
\beq
\label{B}
N=\frac{1}{2}\Bigg(x_1 \gamma_1 \frac{\partial \zeta_1^*}{\partial \gamma_1}
+x_2 \gamma_2 \frac{\partial \zeta_2^*}{\partial \gamma_1}\Bigg).
\eeq
In addition, \vicenteg{in Eqs.\ \eqref{2.9}, \eqref{2.10}, \eqref{a2}, and \eqref{a3}}
\beq
\label{a14}
\left(\frac{\partial \zeta_i^*}{\partial \Delta^*}\right)=\left(\frac{\partial \zeta_i^*}{\partial \Delta^*}\right)_{\gamma_1}+\left(\frac{\partial \zeta_i^*}{\partial \gamma_1}\right)\left(\frac{\partial \gamma_1}{\partial \Delta^*}\right),
\eeq
and
\beq
\label{a15}
\left(\frac{\partial \zeta_i^*}{\partial x_1}\right)=\left(\frac{\partial \zeta_i^*}{\partial x_1}\right)_{\gamma_1}+\left(\frac{\partial \zeta_i^*}{\partial \gamma_1}\right)\left(\frac{\partial \gamma_1}{\partial x_1}\right).
\eeq
Finally, the derivative $\partial \gamma_1/\partial x_1$ can be written as
\begin{widetext}
\beq
\label{a16}
\frac{\partial \gamma_1}{\partial x_1}=-\frac{\gamma_{1}\frac{\partial \zeta_1^*}{\partial x_1}+\frac{1}{2}\left(x_1\gamma_{1}\frac{\partial \zeta_1^*}{\partial x_1}+x_2\gamma_{2}\frac{\partial \zeta_2^*}{\partial x_1}\right)\left[\Delta^* \left(\frac{\partial \gamma_1}{\partial \Delta^*}\right)-2\gamma_1\right]}
{\gamma_{1}\frac{\partial \zeta_1^*}{\partial \gamma_1}+\frac{1}{2}\left(x_1\gamma_{1}\frac{\partial \zeta_1^*}{\partial \gamma_1}+x_2\gamma_{2}\frac{\partial \zeta_2^*}{\partial \gamma_1}\right)\left[\Delta^* \left(\frac{\partial \gamma_1}{\partial \Delta^*}\right)-2\gamma_1\right]}.
\eeq
\end{widetext}
In Eq.\ \eqref{a16}, the derivative $\partial_{x_1}\zeta_i^*$ is taken at $\gamma\equiv \text{const}$.

\section{Eigenvalues in the limit $k\to 0$}
\label{appB}

In this Appendix, we provide the expressions of the eigenvalues of the matrix $\mathsf{M}$ for small wave numbers. In this limit ($k\to 0$), the solution is given by Eq.\ \eqref{2.14.1}. Substitution of the expansion \eqref{2.14.1} into the quartic equation \eqref{2.11} allows one to get the coefficients $s_n^{(\ell)}$.
In the zeroth-order in $k$, one simply achieves the result $s_1^{(0)}=B<0$, $s_2^{(0)}=s_3^{(0)}=s_4^{(0)}=0$. In the first-order in $k$, one gets the result
\beq
\label{b1}
s_n^{(0)2}s_n^{(1)}\left(4s_n^{(0)}-3B\right)=0.
\eeq
When $s_1^{(0)}=B$, the solution to Eq.\ \eqref{b1} is $s_1^{(1)}=0$. However, when $s_n^{(0)}=0$, ($n=2,3,4$), Eq.\ \eqref{b1} applies for any value of $s_n^{(1)}$.

In the second-order in $k$, one gets a relatively long equation involving $s_n^{(1)}$ and $s_n^{(2)}$. When $s_n^{(0)}=0$, this equation applies for any value of $s_n^{(1)}$ and $s_n^{(2)}$. However, when $s_1^{(0)}=B$, $s_1^{(1)}=0$, and
\beq
\label{b2}
s_1^{(2)}=\frac{B\left(M_{22}^{(2)}+M_{23}^{(2)}\right)-M_{24}^{(1)}M_{43}^{(1)}-A\left(M_{12}^{(2)}+M_{13}^{(2)}\right)}{B},
\eeq
where
\begin{widetext}
\beq
\label{5}
M_{22}^{(2)}=\frac{1-\mu}{4\mu_{12}}D_T^{\ast}-\frac{2}{d}\lambda^*, \quad M_{23}^{(2)}=\frac{1-\mu}{4\mu_{12}}D_p^{\ast}-\frac{2}{d}L^*, \quad
M_{12}^{(2)}=-\frac{1}{4x_1}\frac{\mu x_{1}+x_{2}}{\mu_{12}}D_T^*,
\eeq
\beq
\label{6}
M_{13}^{(2)}=-\frac{1}{4x_1}\frac{\mu x_{1}+x_{2}}{\mu_{12}}D_p^*, \quad M_{24}^{(1)}=-\left(\frac{2}{d}+\zeta_U\right), \quad M_{43}^{(1)}=-\frac{1}{4}\frac{1+\mu}{x_1\mu+x_2}.
\eeq
%If the homogeneous steady state (HSS) is stable, then $s_1^{(2)}$ must be negative.

To determine $s_n^{(1)}$ and $s_n^{(2)}$ when $s_n^{(0)}=0$, one needs to consider the expansion of the eigenvalues $s_n$ up to $k^3$ and $k^4$. Thus, in the third-order in $k$, when $s_n^{(0)}=0$ the solutions for $s_n^{(1)}$ ($n=2,3,4$) are
\beq
\label{7}
s_2^{(1)}=0, \quad s_{3,4}^{(1)}=\pm \sqrt{M_{43}^{(1)}\left(M_{24}^{(1)}-M_{34}^{(1)}\right)},
\eeq
where
\beq
\label{7.0}
M_{34}^{(1)}=-\left(\frac{d+2}{d}+\zeta_U\right).
\eeq
According to the expressions of $M_{43}^{(1)}$, $M_{24}^{(1)}$, and $M_{34}^{(1)}$, one achieves the result $s_{3,4}^{(1)}=\pm \imath c$, where
\beq
\label{7.1}
c=\sqrt{\frac{1}{4}\frac{1+\mu}{x_1\mu+x_2}}>0.
\eeq

The remaining coefficients are obtained when one considers the fourth-order in $k$. When $s_2^{(1)}=0$, one gets
\beq
\label{8}
s_2^{(2)}=\frac{B M_{11}^{(2)}+A M_{12}^{(2)}}{B},
\eeq
where
\beq
\label{9}
M_{11}^{(2)}=-\frac{1}{4}\frac{\mu x_{1}+x_{2}}{\mu_{12}}D^*.
\eeq
Since $A<0$, $B<0$, $D^*>0$ , and $D_T^*>0$, then $s_2^{(2)}<0$. On the other hand, when $s_{3,4}^{(1)}=\pm \imath c$, $s_{3,4}^{(2)}$ is given by
\beq
\label{10}
s_{3,4}^{(2)}=\frac{\Gamma}{B\left[\left(M_{24}^{(1)}-M_{34}^{(1)}\right)M_{43}^{(1)}-3 s_{3,4}^{(1)2}\right]},
\eeq
where
\beqa
\label{11}
\Gamma&=&A\left[M_{43}^{(1)}M_{12}^{(2)}\left(M_{24}^{(1)}-M_{34}^{(1)}\right)-s_{3,4}^{(1)2}\left(M_{12}^{(2)}+
M_{13}^{(2)}\right)\right]-s_{3,4}^{(1)2}\left(s_{3,4}^{(1)2}+M_{43}^{(1)}M_{34}^{(1)}\right) \nonumber\\
& & +B\left[M_{43}^{(1)}M_{11}^{(2)}\left(M_{24}^{(1)}-M_{34}^{(1)}\right)-s_{3,4}^{(1)2}\left(M_{11}^{(2)}-
M_{23}^{(2)}+M_{33}^{(2)}+M_{44}^{(2)}\right)\right].
\eeqa
Here, we have introduced the quantities
\beq
\label{12}
M_{23}^{(2)}=\frac{1-\mu}{4 \mu_{12}}D_p^*-
\frac{2}{d}L^{\ast}, \quad M_{33}^{(2)}=-\frac{2}{d}L^*,
\quad M_{44}^{(2)}=-\frac{1+\mu}{\mu x_1+x_2}\frac{d-1}{d}\eta^*.
\eeq
%As said before, if the HSS is linearly stable, then $s_{3,4}^{(2)}<0$.
\end{widetext}

%\bibliography{diffusionDelta}

\begin{thebibliography}{65}
\bibitem{note}\vicenteg{A misprint in Ref.\ \onlinecite{GBS21} for the expression of $Z$ was found while the present paper was written. The expression \eqref{a11.1} for the first term in $Z$ corrects this mistake. This typo affects some of the graphs displayed in Ref.\ \onlinecite{GBS21}. The correct Mathematica code for computing the dependence of the set of transport coefficients on the parameter space of the system is available to the interested readers from the authors upon request.}
\end{thebibliography}

%\end{document}

%

\end{document}